\documentclass[twoside,slac_one]{revtex4}
\usepackage{graphicx}
\usepackage{fancyhdr}
\usepackage{amsmath} 
\usepackage{bm}
\usepackage{amsxtra}
\usepackage{amssymb}
\usepackage{amsthm}
\usepackage{latexsym}
\usepackage{lscape}

\begin{document}

\title{Frequency Scanned Interferometry for ILC Tracker Alignment}

%

\author{Hai-Jun Yang$^1$, Tianxiang Chen$^{1*}$, Keith Riles$^1$}
\affiliation{
$^1$ Department of Physics, University of Michigan, Ann Arbor, MI 48109-1120, USA \\
$^*$ Department of Physics, University of Science and Technology of China, Hefei, China}

\begin{abstract}
In this paper, we report high-precision absolute distance and vibration
measurements performed with frequency scanned interferometry.
Absolute distance was determined 
by counting the interference fringes produced while scanning the laser frequency. 
High-finesse Fabry-Perot interferometers were used to determine frequency 
changes during scanning. A dual-laser scanning technique 
was used to cancel drift errors to improve the absolute distance measurement precision. 
A new dual-channel FSI demonstration system is also presented which is an interim stage toward
practical application of multi-channel distance measurement.
Under realistic conditions, a precision of 0.3 microns was achieved for an
absolute distance of 0.57 meters. 
A possible optical alignment system for a silicon tracker is also presented.
\end{abstract}

\maketitle

\thispagestyle{fancy}


\section{Introduction}
The motivation for this project is to design a novel optical
system for quasi-real time alignment of tracker detector elements
used in High Energy Physics (HEP) experiments. A.F. Fox-Murphy {\em et.al.}
from Oxford University reported their design of a frequency
scanned interferometer (FSI) for precise alignment of the ATLAS
Inner Detector \cite{ox98,ox04,ox05}. Given the demonstrated need for
improvements in detector performance, we plan to design and prototype an
enhanced FSI system to be used for the alignment of tracker
elements in the next generation of electron-positron Linear
Collider detectors. Current plans for future detectors
require a spatial resolution for signals from a tracker detector,
such as a silicon microstrip or silicon drift detector,
to be approximately 7-10 $\mu m$\cite{orangebook}. To achieve this
required spatial resolution, the measurement precision of absolute
distance changes of tracker elements in one dimension should be on the order of 1
$\mu m$. Simultaneous measurements from hundreds of interferometers
will be used to determine the 3-dimensional positions of the
tracker elements.

In this paper, we describe ongoing R\&D in frequency scanned interferometry (FSI) 
to be applied to alignment monitoring of a detector's charged particle tracking system, 
in addition to its beam pipe and final-focus quadrupole magnets. 

The University of Michigan group has constructed several demonstration FSIs with the laser
light transported by air or single-mode optical fiber, using single-fiber and dual-laser
scanning techniques, and dual-laser with dual-channel for initial feasibility studies.
Absolute distance was determined by counting the interference
fringes produced while scanning the laser frequency.
The main goal of the demonstration systems was to determine the
potential accuracy of absolute distance 
measurements that could be achieved under both controlled and realistic conditions.
Secondary goals included estimating the effects of vibrations and studying
error sources crucial to the absolute distance accuracy.
Two multiple-distance-measurement analysis techniques
were developed to improve distance precision and to extract the amplitude
and frequency of vibrations.
Under laboratory conditions, a measurement precision of $\sim$ 50 nm was achieved for
absolute distances ranging from 0.1 meters to 0.7 meters by using the first
multiple-distance-measurement technique.
The second analysis technique has the capability to
measure vibration frequencies ranging from 0.1 Hz to 100 Hz with amplitude
as small as a few nanometers, without a {\em priori} knowledge\cite{fsi05}. 
The multiple-distance-measurement analysis techniques are well suited for
reducing vibration effects and uncertainties from fringe \& frequency determination,
but do not handle well the drift errors, such as from thermal effects.

We describe a dual-laser system intended to reduce the drift errors
and show some results under realistic conditions. The dual-channel FSI is used to
make sanity checks of the displacement of the detector simultaneously.
Dual lasers with oppositely scanned frequency directions permit cancellation of many 
systematic errors, making the alignment robust against vibrations and environmental 
disturbances. 

We also report on progress using a dual-channel dual-laser FSI with prototype. 
Under realistic environmental conditions, a precision of about 0.2-0.3 microns was achieved 
for a distance of about 57 cm for the prototype.

\section{Principles}
The intensity $I$ of any two-beam interferometer can be expressed as
$I = I_1 + I_2 + 2\sqrt{I_1 I_2} \cos(\phi_1 - \phi_2)$, 
where $I_1$ and $I_2$ are the intensities of the two combined beams,
and $\phi_1$ and $\phi_2$ are the phases.
Assuming the optical path lengths of the two beams are $D_1$ and
$D_2$, the phase difference is
$\Phi = \phi_1 - \phi_2 = 2\pi |D_1 - D_2|(\nu/c)$,
where $\nu$ is the optical frequency of the light, and c is
the speed of light.

For a fixed path interferometer, as the frequency of the laser is
continuously scanned, the optical beams will
constructively and destructively interfere, causing ``fringes''.
The number of fringes $\Delta N$ is $\Delta N = D\Delta\nu/c$,
where $D$ is the optical path difference between the two beams,
and $\Delta\nu$ is the scanned frequency range. The optical path
difference (OPD for absolute distance between beamsplitter and retroreflector)
can be determined by counting interference fringes while scanning the laser frequency.

\begin{figure}[htbp]
\includegraphics[width=80mm]{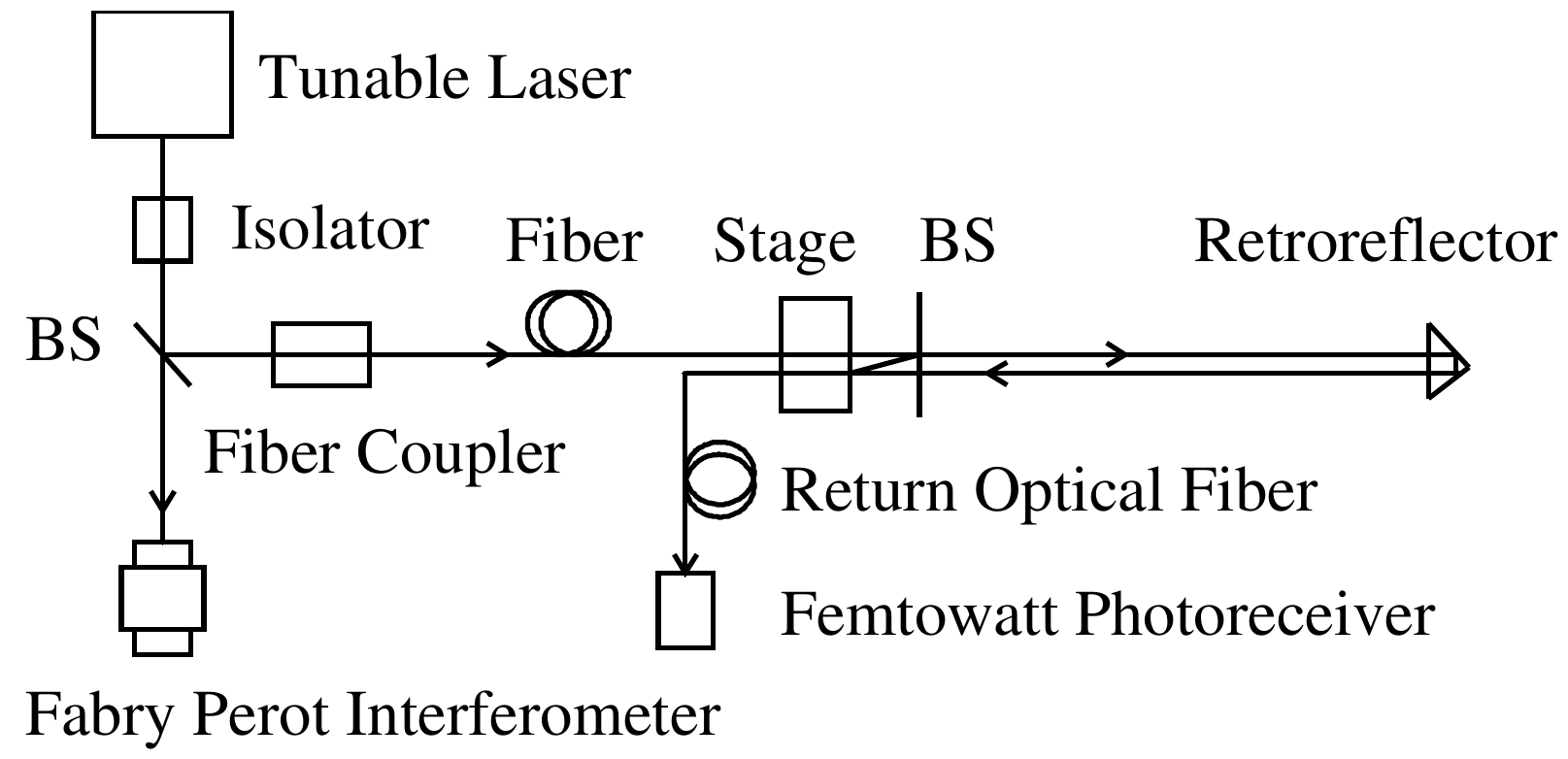}
\caption{\label{fsi} Schematic of an optical fiber FSI system.}
\end{figure}

If small vibration and drift errors $\epsilon(t)$ occur during the laser scanning, then
$\Phi(t) =  2\pi \times (D_{true}+\epsilon(t)) \times \nu(t)/c$, 
$\Delta N = [\Phi(t) - \Phi(t0)]/2\pi = 
D_{true}\Delta\nu/c  + [\epsilon(t)\nu(t)/c - \epsilon(t0)\nu(t0)/c]$, 
Assuming $\nu(t) \sim \nu(t0) = \nu$, $\Omega = \nu/\Delta\nu$, 
$\Delta \epsilon = \epsilon(t) - \epsilon(t0)$,
the measured distance can be written as,
\begin{eqnarray}
D_{measured} = \Delta N / (\Delta\nu/c) = D_{true} + \Delta\epsilon \times \Omega.
\label{eq_fsi}
\end{eqnarray}

\section{Demonstration System of FSI}

A schematic of the FSI system with a pair of optical fibers is shown in Figure~\ref{fsi}
The light source is a New Focus Velocity 6308 tunable laser 
(665.1 nm $<\lambda<$ 675.2 nm). A high-finesse ($>200$) Thorlabs SA200
F-P is used to measure the frequency range scanned by the laser. The free 
spectral range (FSR) of two adjacent F-P peaks is 1.5 GHz, which corresponds 
to 0.002 nm. A Faraday Isolator was used to reject light reflected back into
the lasing cavity. The laser beam was coupled into a single-mode optical fiber
with a fiber coupler.
Data acquisition is based on a National Instruments DAQ
card capable of simultaneously sampling 8 channels at a rate of 250 KS/s/ch with
a precision of 12-bits.  Omega thermistors with a tolerance of 0.02 K and a
precision of 0.01 $mK$ are used to monitor temperature.  The apparatus
is supported on a damped Newport optical table.

In order to reduce air flow and temperature fluctuations, a transparent plastic
box was constructed on top of the optical table. PVC pipes were installed to 
shield the volume of air surrounding the laser beam. Inside the PVC pipes, 
the typical standard deviation of 20 temperature measurements was about $0.5 ~mK$. 
Temperature fluctuations were suppressed by a factor of approximately 100 by 
employing the plastic box and PVC pipes.

Detectors for HEP experiments must usually be operated remotely for 
safety reasons because of intensive radiation, high voltage or strong magnetic
fields. In addition, precise tracking elements are typically surrounded by
other detector components, making access difficult. For practical HEP application
of FSI, optical fibers for light delivery and return are therefore necessary.

The beam intensity coupled into the return optical fiber is very weak, requiring
ultra-sensitive photodetectors for detection. Considering the limited laser beam 
intensity used here and the need to split into many beams to serve a set
of interferometers, it is vital to increase the geometrical efficiency.
To this end, a collimator is attached to the optical fiber, the density of 
the outgoing beam from the optical fiber is increased significantly.
The return beams are received by another optical fiber and amplified by a 
Si femtowatt photoreceiver with a gain of $2 \times 10^{10} V/A$. 

\section{Multiple-Distance-Measurement Techniques}

For a FSI system, drifts and vibrations occurring along the optical
path during the scan will be magnified by a factor of $\Omega =
\nu / \Delta \nu$, where $\nu$ is the average optical frequency of the laser
beam and $\Delta \nu$ is the scanned frequency range. For the full scan of
our laser, $\Omega \sim 67 $. Small vibrations and
drift errors that have negligible effects for many optical
applications may have a significant impact on a FSI system.
A single-frequency vibration
may be expressed as $x_{vib}(t) = a_{vib} \cos(2\pi f_{vib} t + \phi_{vib})$,
where $a_{vib}$, $f_{vib}$ and $\phi_{vib}$ are the amplitude, frequency and phase of the
vibration, respectively.
If $t_0$ is the start time of the scan, Eq.~\ref{eq:rdist} can be re-written as
\begin{eqnarray}
\Delta N=L\Delta\nu/c+2[x_{vib}(t)\nu(t)-x_{vib}(t_0)\nu(t_0)]/c
\label{eq:rdist}
\end{eqnarray}

If we approximate $\nu(t) \sim \nu(t_0) = \nu$, the
measured optical path difference $L_{meas}$ may be expressed as

\begin{eqnarray}
L_{meas} = L_{true} - 4 a_{vib} \Omega \sin[\pi f_{vib}(t-t_0)] \times \sin[\pi f_{vib}(t+t_0)+\phi_{vib}]
\label{eq:rdist_vib}
\end{eqnarray}

where $L_{true}$ is the true optical path difference in the absence of 
vibrations. If the path-averaged refractive index of
ambient air $\bar{n}_g$ is known, the measured distance
is $R_{meas} = L_{meas}/(2\bar{n}_g)$.

\begin{figure}[htbp]
\includegraphics[width=11cm]{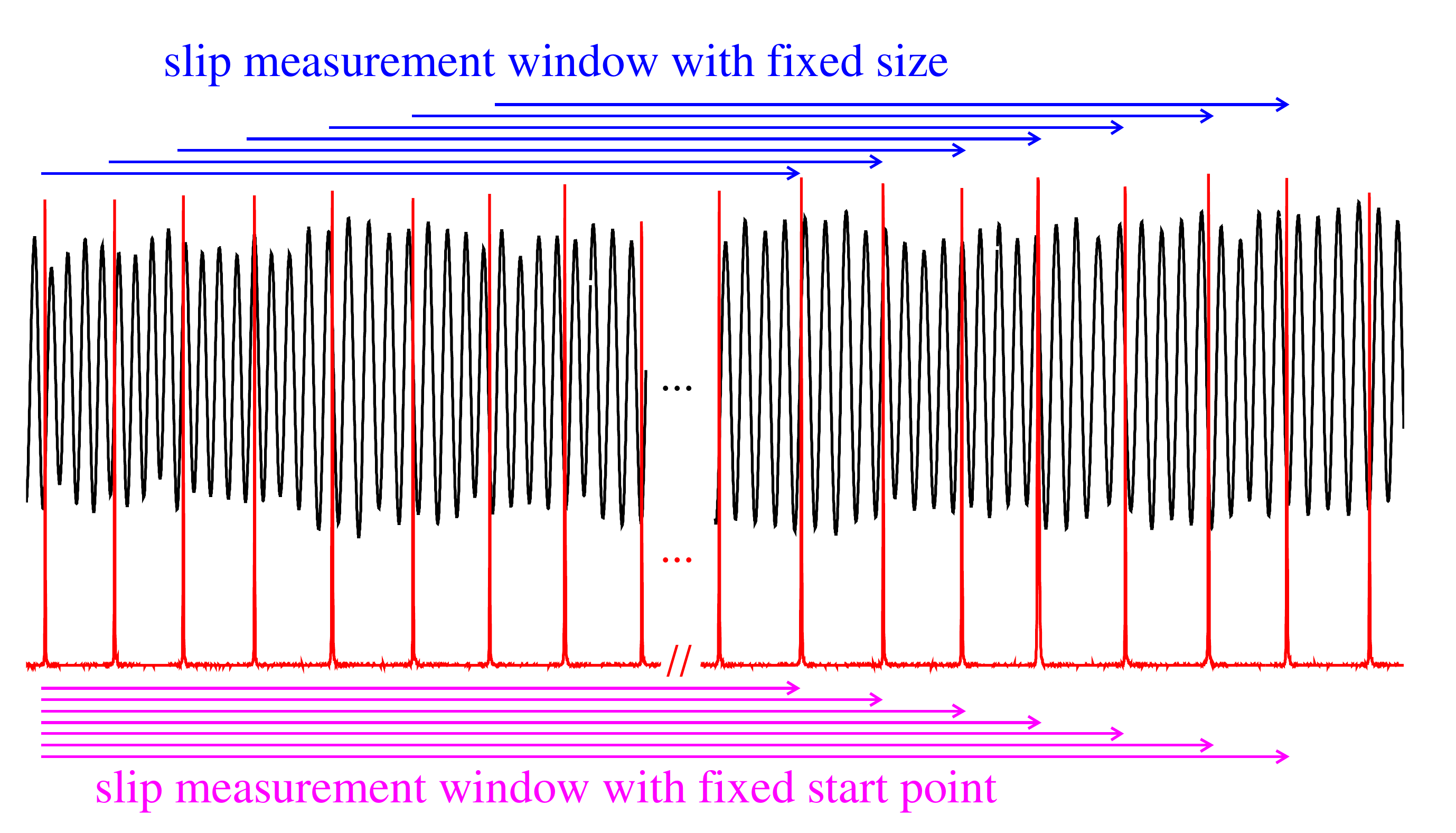}
\caption{\label{multi_dist} The schematic of two multiple-distance-measurement techniques.
The interference fringes from the femtowatt photoreceiver and
the scanning frequency peaks from the Fabry-Perot interferometer(F-P)
for the optical fiber FSI system recorded simultaneously by DAQ card
are shown in black and red, respectively. The free spectral range(FSR)
of two adjacent F-P peaks (1.5 GHz) provides a calibration of the scanned frequency range.}
\end{figure}

If the measurement window size $(t-t_0)$ is fixed and the window used to
measure a set of $R_{meas}$ is sequentially shifted, the
effects of the vibration will be evident. 
We use a set of distance measurements in one scan by successively shifting the 
fixed-length measurement window one F-P peak forward each time. 
The arithmetic average of all measured $R_{meas}$ values in one scan is taken to be the
measured distance of the scan (although more sophisticated fitting methods
can be used to extract the central value). 
For a large number of
distance measurements $N_{meas}$, the vibration effects can be greatly suppressed. 
Of course, statistical uncertainties
from fringe and frequency determination, dominant in
our current system, can also be reduced with multiple scans.
Averaging multiple measurements in one scan, however, provides similar precision
improvement to averaging distance measurements from 
independent scans, and is faster, more efficient, and less 
susceptible to systematic errors from drift.
In this way, we can improve the distance accuracy dramatically if 
there are no significant drift errors during one scan, caused, for example, 
by temperature variation. 
This multiple-distance-measurement technique is called 'slip measurement
window with fixed size', shown in Figure~\ref{multi_dist}. However, there is a trade off in that the
thermal drift error is increased with the increase of $N_{meas}$
because of the larger magnification factor $\Omega$ for a smaller
measurement window size.

In order to extract the amplitude and frequency of the vibration, another 
multiple-distance-measurement technique called 'slip measurement window with 
fixed start point' is used, as shown in Figure~\ref{multi_dist}. 
In Eq.~\ref{eq:rdist_vib}, if $t_0$ is fixed, the 
measurement window size is enlarged one F-P peak for each shift, an 
oscillation of a set of measured $R_{meas}$ values indicates the amplitude 
and frequency of vibration. 
This technique is not suitable for distance measurement because there
always exists an initial bias term, from $t_0$,
which cannot be determined accurately in our current system.

\section{Absolute Distance and Vibration Measurement}

\begin{figure}[htbp]
\includegraphics[width=65mm]{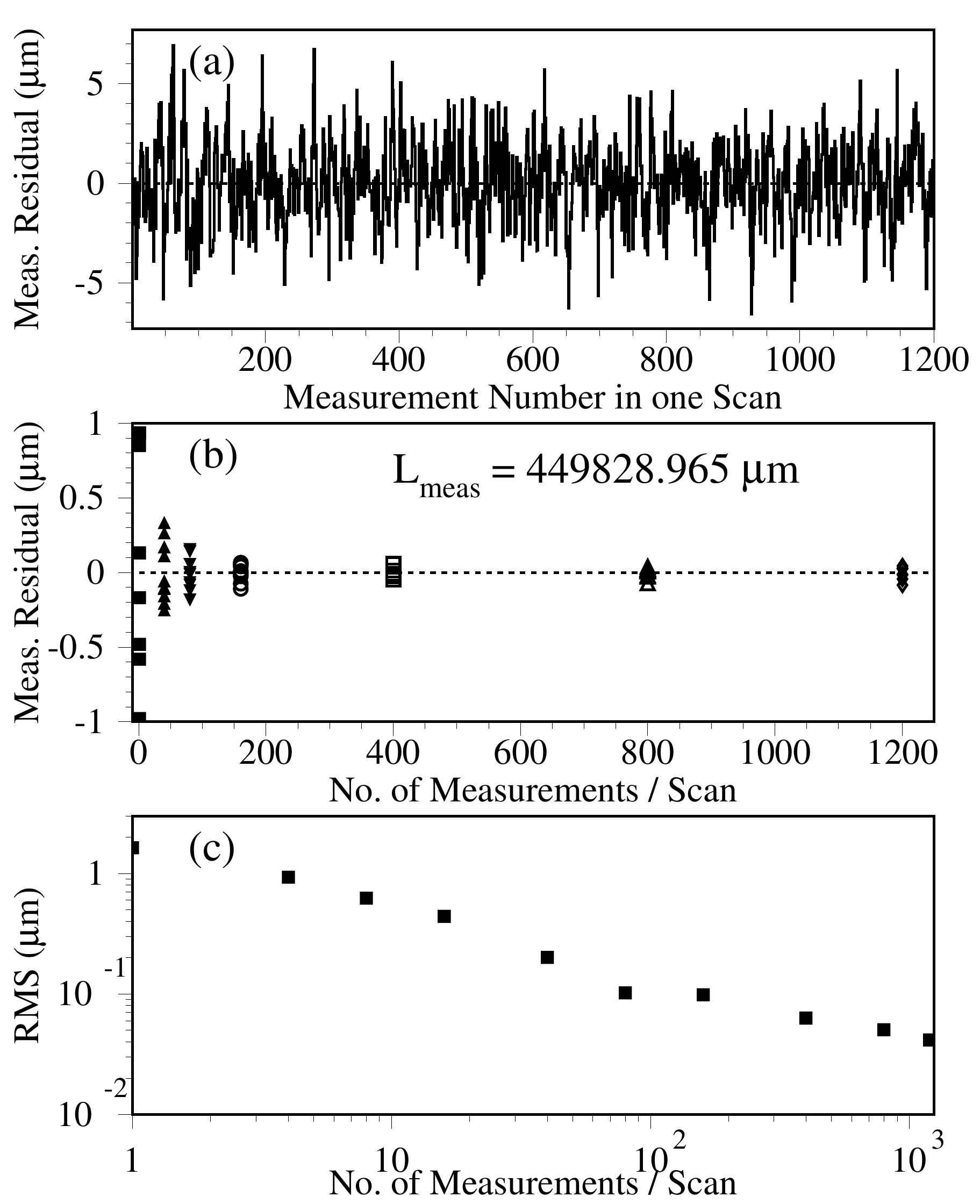}
\includegraphics[width=70mm]{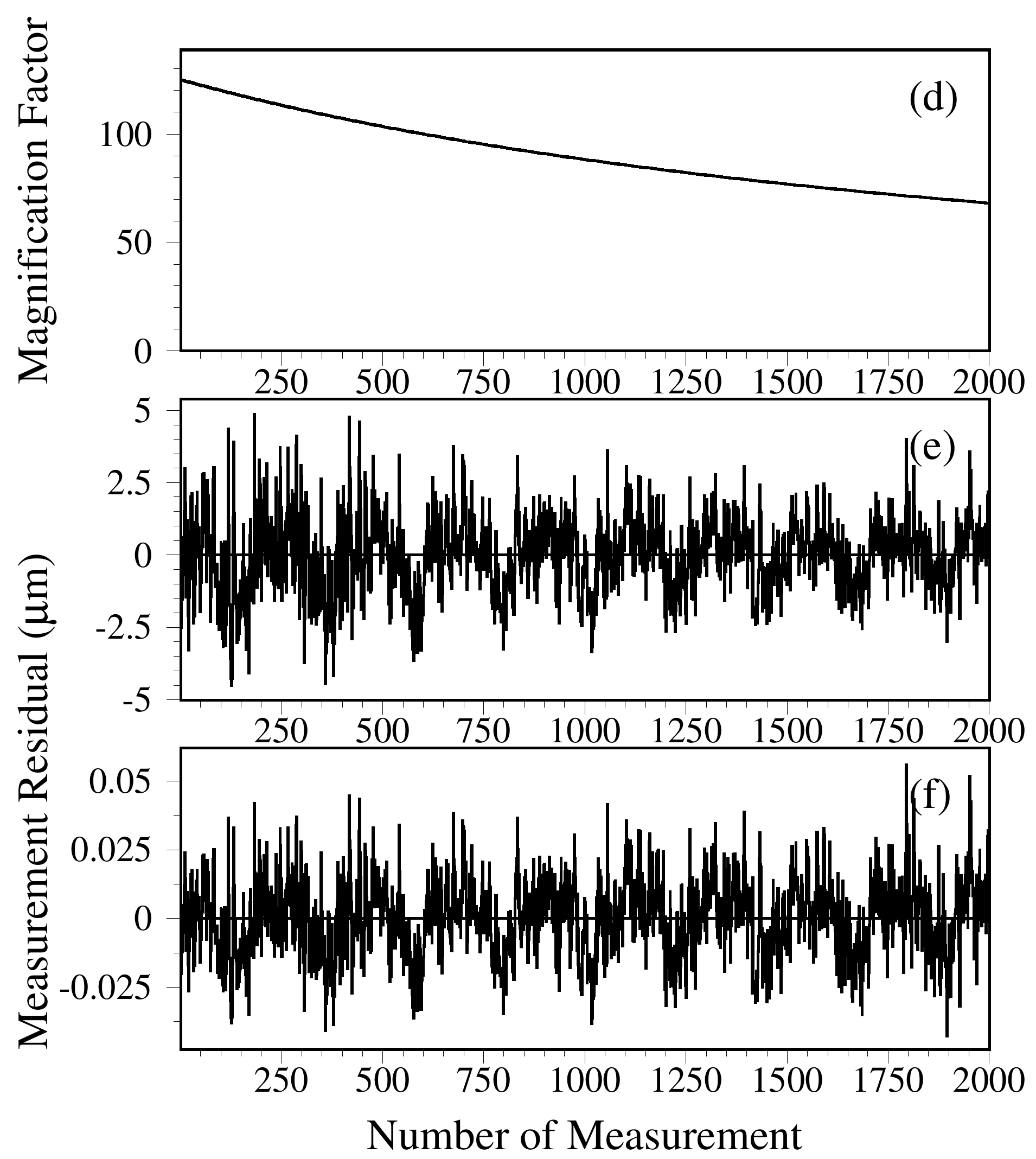}
\caption{\label{rdist} Distance measurement residual spreads versus number of distance measurement $N_{meas}$ 
(a) for one typical scan, 
(b) for 10 sequential scans,
(c) is the standard deviation of distance measurements for 10 sequential scans
versus $N_{meas}$.
The frequency and amplitude of the controlled vibration source 
are 1 Hz and 9.5 nanometers, 
(d) Magnification factor  versus number of distance measurements,
(e) Distance measurement residual  versus number of distance measurements,
(f) Corrected measurement residual  versus number of distance measurements.}
\end{figure}

The typical measurement residual versus the distance
measurement number in one scan using the above technique is shown in Figure~\ref{rdist}(a), where
the scanning rate was 0.5 nm/s and the sampling rate was 125 kS/s.
Measured distances minus their average value for 10 sequential scans
are plotted versus number of measurements ($N_{meas}$) per scan in Figure~\ref{rdist}(b).
The standard deviations (RMS) of distance measurements for 
10 sequential scans are plotted versus number of measurements ($N_{meas}$) per scan in Figure~\ref{rdist}(c).
It can be seen that the distance errors decrease with an increase of $N_{meas}$.
The RMS of measured distances for 10 sequential scans
is 1.6 $\mu m$ if there is only one distance measurement per scan ($N_{meas}=1$).
If $N_{meas}=1200$ and the average value of 1200 distance measurements
in each scan is considered as the final measured distance of the scan,
the RMS of the final measured distances for 10 scans is 41 nm for the distance of
449828.965 $\mu m$, the relative distance measurement precision is 91 ppb.

The standard deviation (RMS) of measured distances for 10 sequential scans 
is approximately 1.5 $\mu m$ if there is only one distance measurement per scan
for closed box data. By using the multiple-distance-measurement 
technique, the distance measurement precisions for various closed box data
with distances ranging from 10 cm to 70 cm collected 
are improved significantly; precisions of approximately 50 nanometers are
demonstrated under laboratory conditions, as shown in Table 1. 
All measured precisions
listed in Table~\ref{tab:dist_single_laser} are the RMS's of measured distances for 10 sequential scans.
Two FSI demonstration systems, 'air FSI' and 'optical fiber FSI',
were constructed for extensive tests of multiple-distance-measurement technique, 
'air FSI' means FSI with the laser beam transported entirely in the ambient atmosphere,
'optical fiber FSI' represents FSI with the laser beam delivered to the interferometer 
and received back by single-mode optical fibers. 

\begin{table}
\begin{tabular}{|c|c|c|c|c|} \hline
Distance & \multicolumn{2}{|c|}{Precision($\mu m$)} & Scanning Rate & FSI System \\ 
\cline{2-3}
(cm) & open box & closed box & (nm/s) & (Optical Fiber or Air) \\ \hline
10.385107 & 1.1 & 0.019 & 2.0 & Optical Fiber FSI \\ \hline
10.385105 & 1.0 & 0.035 & 0.5 & Optical Fiber FSI \\ \hline
20.555075 & - & 0.036, 0.032 & 0.8 & Optical Fiber FSI \\ \hline
20.555071 & - & 0.045, 0.028 & 0.4 & Optical Fiber FSI \\ \hline
41.025870 & 4.4 & 0.056, 0.053 & 0.4 & Optical Fiber FSI \\ \hline
44.982897 & - & 0.041 & 0.5 & Optical Fiber FSI \\ \hline 
61.405952 & - & 0.051 & 0.25 & Optical Fiber FSI \\ \hline
65.557072 & 3.9, 4.7 & - & 0.5 & Air FSI \\ \hline
70.645160 & - & 0.030, 0.034, 0.047 & 0.5 & Air FSI \\ \hline
\end{tabular}
\caption{\label{tab:dist_single_laser} Distance measurement precisions for various setups using the 
multiple-distance-measurement technique.}
\end{table}

Based on our studies, the slow fluctuations are reduced to a negligible level by using 
the plastic box and PVC pipes to suppress temperature fluctuations.
The dominant error comes from the uncertainties of the interference fringes 
number determination; the fringes uncertainties are uncorrelated for multiple 
distance measurements. In this case, averaging multiple distance measurements
in one scan provides a similar precision improvement to averaging distance measurements
from multiple independent scans. But, for open box data, the slow fluctuations 
are dominant, on the order of few microns in our laboratory. The 
measurement precisions for single and multiple distance open-box measurements 
are comparable, which indicates that the slow fluctuations cannot be adequately suppressed by using
the multiple-distance-measurement technique. Improvement using a dual-laser system
will be discussed in the next section.

In order to test the vibration measurement technique, 
a piezoelectric transducer (PZT) was employed to produce 
vibrations of the retroreflector. For instance, the frequency of the controlled 
vibration source was set to $1.01 \pm 0.01$ Hz with amplitude
$9.5 \pm 1.5$ nanometers. The magnification factors,
distance measurement residuals and corrected measurement residuals for 2000
measurements in one scan are shown in Figure~\ref{rdist}(d), (e) and (f), respectively.
The extracted vibration frequencies and amplitudes using this technique,
$f_{vib} = 1.025 \pm 0.002$~Hz, $A_{vib} = 9.3 \pm 0.3 $ nanometers,
agree well with the expectation values.

Detailed information about estimation of major error sources for the absolute distance
measurement and limitation of our current FSI system is provided elsewhere\cite{fsi05}.

\section{Dual-Laser FSI System}

A dual-laser FSI system was built in order to reduce drift error and slow
fluctuations occuring during the laser scan.
Two lasers are operated simultaneously; the two laser beams are coupled into one optical
fiber but temporally isolated by using two choppers.
The principle of the dual-laser technique\cite{ox04} is shown in the following.
For the first laser, the measured distance 
$D_1 = D_{true} + \Omega_1 \times \Delta\epsilon_1$,
and $\Delta\epsilon$ is drift error during the laser scanning.
For the second laser, the measured distance 
$D_2 = D_{true} + \Omega_2 \times \Delta\epsilon_2$.
Since the two laser beams travel the same optical path during the same period,
the drift errors $\Delta\epsilon_1$ and $\Delta\epsilon_2$ should be very comparable.
Under this assumption, the true distance can be extracted using the formula
$D_{true} =  (D_2-\rho \times D_1)/(1-\rho)$, where, $\rho = \Omega_2/\Omega_1$,
the ratio of magnification factors from two lasers. If two identical lasers
scan the same range in opposite directions simultaneously, then $\rho \simeq -1.0$, and $D_{true}$
can be written as,
\begin{eqnarray}
D_{true} =  (D_2-\rho \times D_1)/(1-\rho) \simeq (D_2 + D_1)/2.0
\label{eq:dual_laser}
\end{eqnarray}

\begin{figure}
\includegraphics[width=8cm,angle=-90]{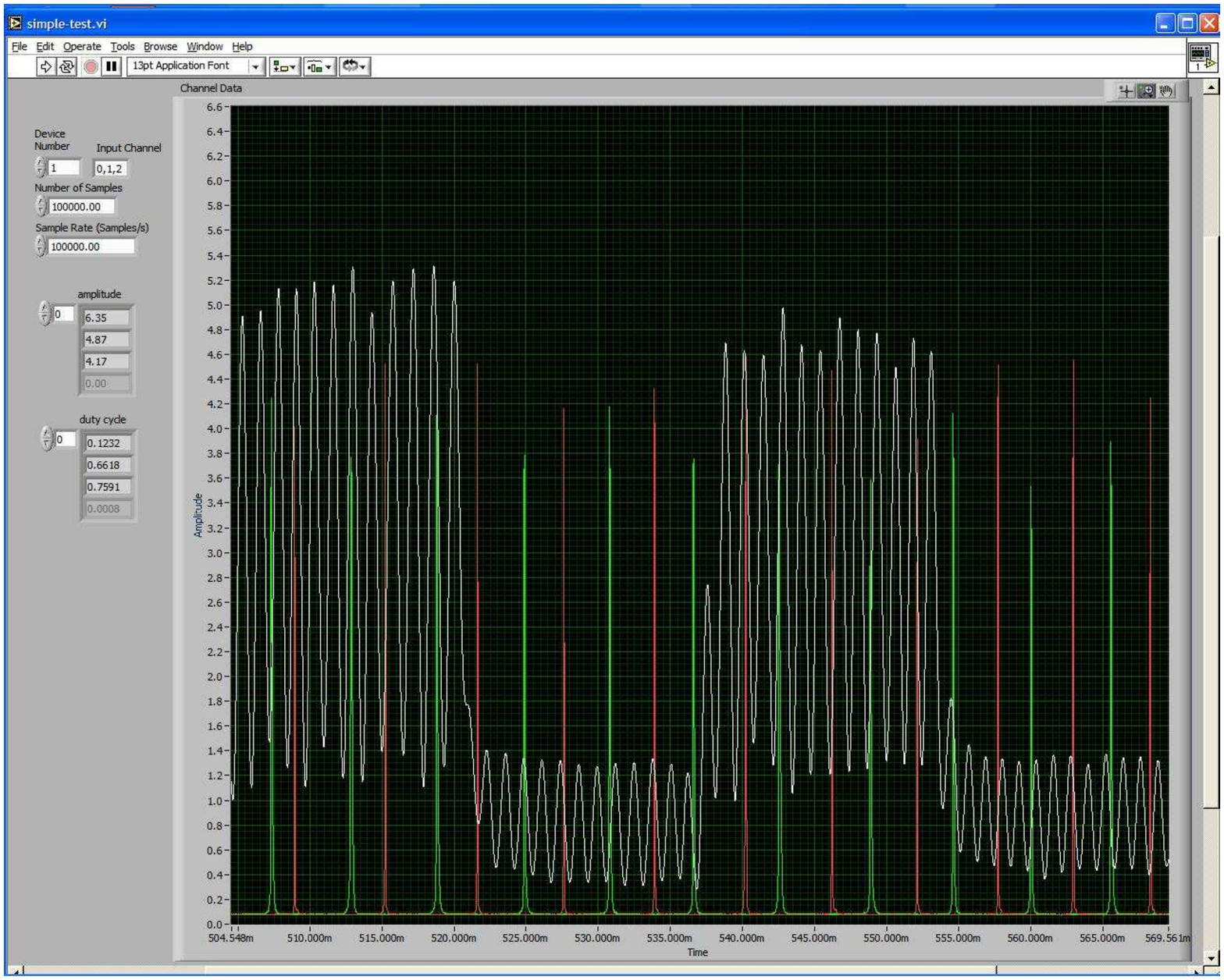}
\includegraphics[width=10cm,angle=0]{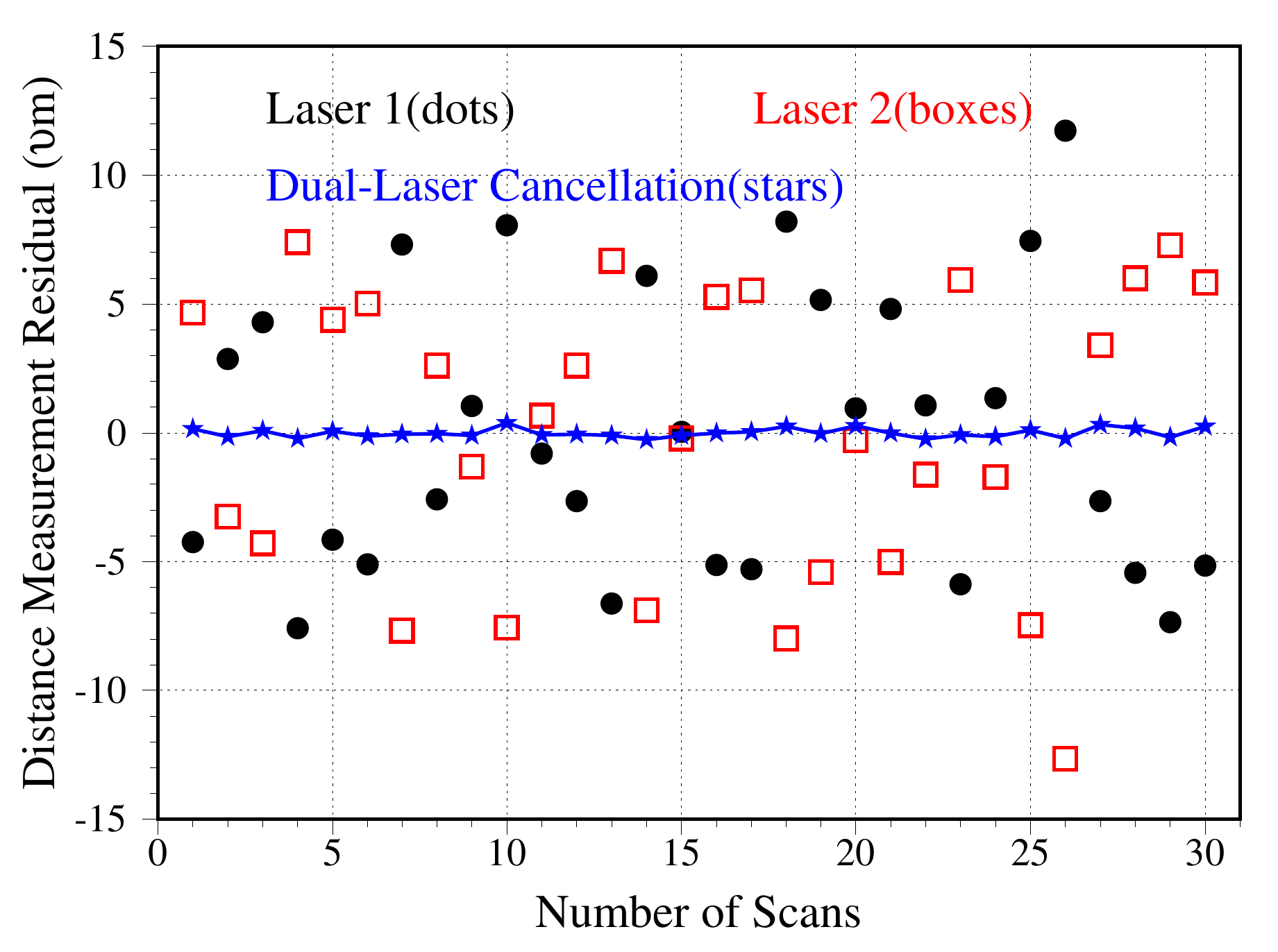}
\caption{\label{dual_laser} The top figure shows interference 
fringes and Fabry-Perot peaks from dual-laser scanning, 
the bottom plot represents distance measurement residual with and without dual-laser cancellation.}
\end{figure}

The laser beams are isolated by choppers periodically, so only half the fringes are
recorded for each laser, degrading the distance measurement precision, as shown in 
the top plot of Figure~\ref{dual_laser}.
Missing fringes during chopped intervals for each laser must be
recovered through robust interpolation algorithms. Based on our studies, the number of
interference fringes in a certain number of Fabry-Perot peaks region is quite stable.
The measured number of fringes is within 0.5 (typically within 0.3)
of the expected fringes count, which enables us to estimate the number of fringes in 
the chopper-off slots(laser beam is blocked by the chopper). In order to determine the
number of fringes in one chopper-off slot, we need to identify two Fabry-Perot peaks within 
two adjacent chopper-on slots closest to the chopper-off slot. If the fringe phases at
the two Fabry-Perot peaks positions are $I+\Delta I$ and $J + \Delta J$, where I and J
are integers, $\Delta I$ and $\Delta J$ are fraction of fringes; then the number
of true fringes can be determined by minimizing the quantity $|N_{correction} + (J+\Delta J) - 
(I + \Delta I) - N_{expected-average}|$, where $N_{correction}$ is an integer
used to correct the fringe number in the chopper off slot, $N_{expected-average}$ is
the expected average number of fringes, based on a full laser 
scanning sample.

Under realistic conditions with large thermal fluctuations, air flow (large drift errors),
10 sequential dual-laser scans data samples each with open box, with a fan on and then
off, were collected. The two lasers were scanned oppositely in frequency over the same band
with scanning speed of 0.4 nm/s, the scanning time is 25 seconds for one full scan.
The measured precision is found to be about $\sim$3-6 microns if 
we use the fringes of these data samples from only one laser
with measured distance of 0.41 meters. If we combine measured distances
from two lasers using Eq.~\ref{eq:dual_laser}, then the dual-laser precision is 0.16 microns for
open box data and 0.20 microns for open box data with the fan on shown in the
bottom plot of Figure~\ref{dual_laser}. Detailed information can be found in
another publication~\cite{fsi07}.

\section{Dual-laser with dual-channel FSI System}

\begin{figure}
\includegraphics[width=10cm,angle=0]{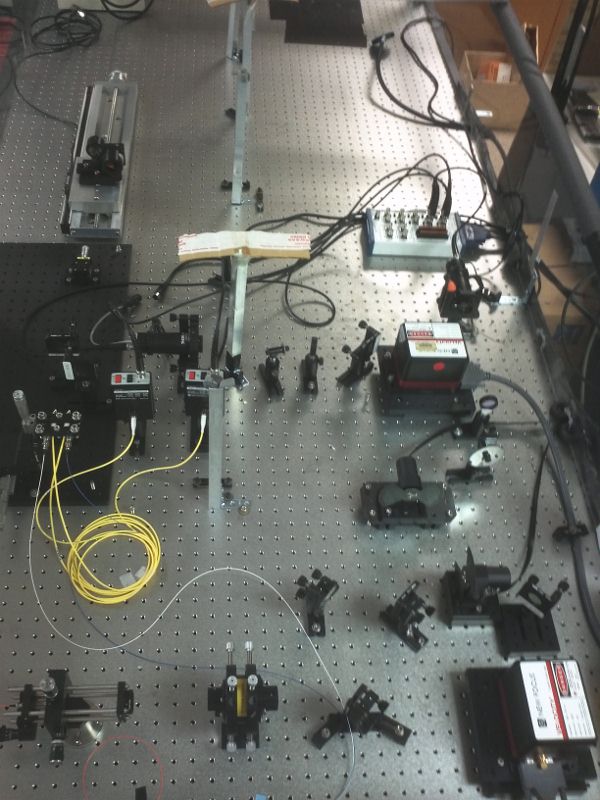}
\caption{\label{dual_channel} The demonstration FSI system with dual-laser and dual-channel.}
\end{figure}

Simultaneous multi-channel distance measurements are required 
for practical application of an FSI system to ILC tracker alignment. 
To this end, we built dual-channel FSI as shown in Figure~\ref{dual_channel}.
The laser beam is coupled into a single mode optical fiber splitter which
split incoming beam into two outgoing beams for two point-to-point FSI
distance measurements. The two retroreflectors are mounted on same tunable 
stage and change positions simultaneously, allowing 
cross check of the displacement of the retroreflectors (detector). 
The results for both FSI channels using dual-laser scanning technique 
are listed in Table~\ref{tab:dist_dual_ch_dual_laser},
where the measured point-to-point distance is about 57 centermeters,
and the tuning stage changes the positions of two retroreflectors by $20~\pm~2$ microns
along the beam line. As can be seen, the measured distance change from two FSI channels
are consistent at each position change, the displacement measurement precision is 
about $0.2-0.3$ microns.

\begin{table}
\begin{tabular}{|c|c|c|} \hline
Distance          & \multicolumn{2}{|c|}{Dual-Laser} \\ \cline{2-3}
Change ($\mu m$)  & Channel 1  & Channel 2         \\ \hline
d2-d1       & 20.746$\pm$0.353 & 21.175$\pm$0.338  \\ \hline
d3-d2       & 20.342$\pm$0.222 & 20.604$\pm$0.239  \\ \hline
d4-d3       & 20.020$\pm$0.219 & 19.878$\pm$0.195  \\ \hline
d5-d4       & 19.968$\pm$0.227 & 20.276$\pm$0.184  \\ \hline
\end{tabular}
\caption{\label{tab:dist_dual_ch_dual_laser} 
Measured distance changes for various setups dual-laser and dual-channel technique.}
\end{table}

\section{A Possible Silicon Tracker Alignment System}

\begin{figure}[htbp]
\includegraphics[width=7cm]{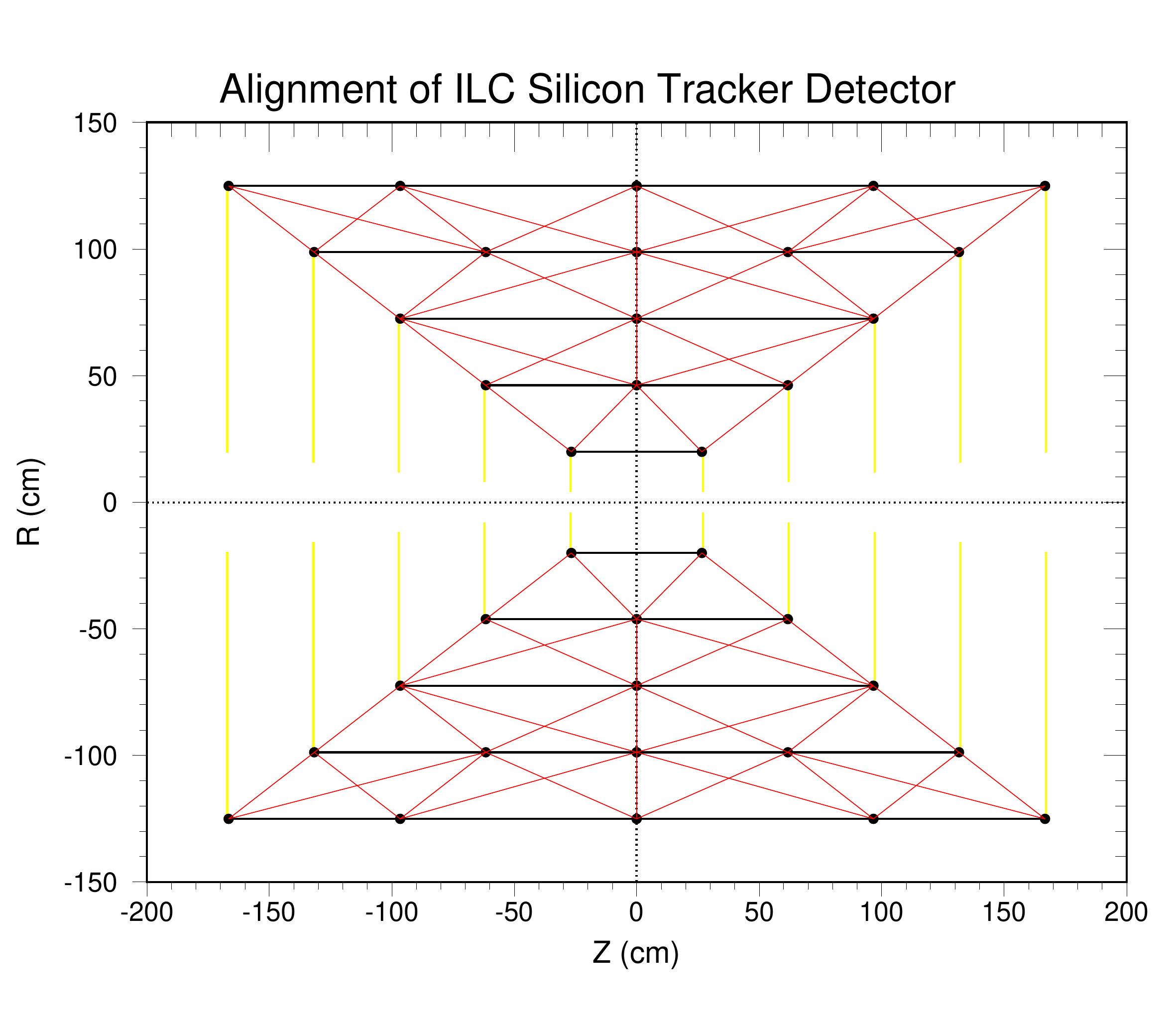}
\includegraphics[width=7cm]{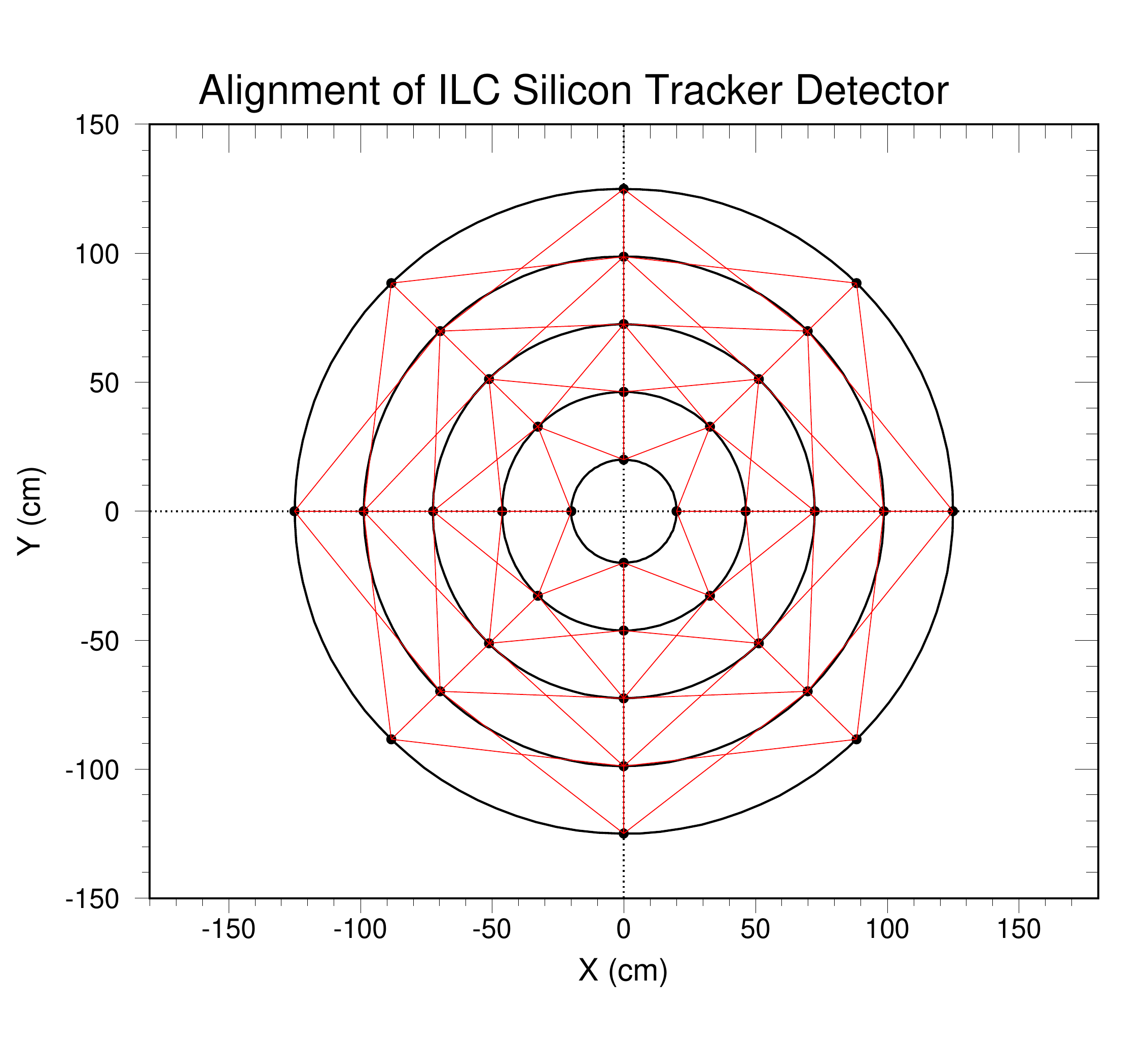}
\includegraphics[width=8cm]{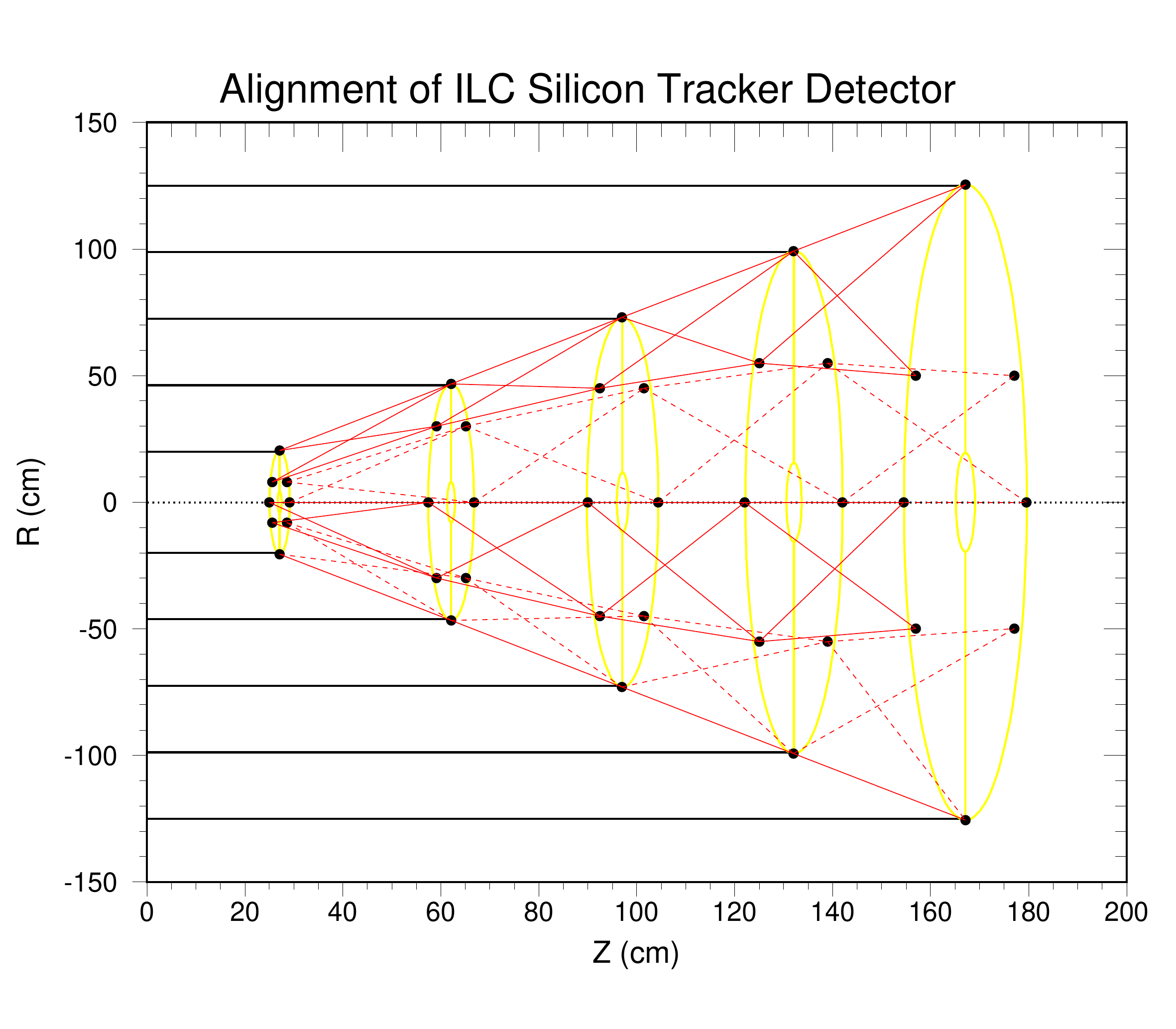}
\caption{\label{sid_alignment} A Possible SiLC Tracker Alignment System.}
\end{figure}

One possible silicon tracker alignment system is shown in Figure~\ref{sid_alignment} 
for a generic tracker.
The top left plot shows lines of sight for alignment in the R-Z plane of the tracker barrel,
the top right plot for alignment in X-Y plane of the tracker barrel,
the bottom plot for alignment in the tracker forward region.
Red lines/dots show the point-to-point distances need to be
measured using FSIs. There are 752 point-to-point distance
measurements in total for the alignment system. More studies
are needed to optimize the distance measurements grid.

%





\begin{acknowledgments}
This work is supported by the National Science Foundation and the Department of Energy of
the United States.
\end{acknowledgments}

\bigskip 

\end{document}